%% file: main.tex
\documentclass[fleqn,11pt]{wlscirep}
\usepackage[T1]{fontenc}
\usepackage[utf8]{inputenc}
\usepackage{upgreek}
\usepackage{graphicx}
\makeatletter
\DeclareTextCommand{\textprime}{\encodingdefault}{%
  \mbox{$\m@th'\kern-\scriptspace$}%
}
\makeatother

\title{A self-referenced in-situ arrival time monitor for X-ray free-electron lasers}

\author[1,2,*]{Michael~Diez}        % badge no. 0039554
\author[1,2]{Andreas~Galler}        % badge no. 0031635
\author[1,2,a,*]{Sebastian~Schulz}  % badge no. 0043229
\author[1,a]{Christina~Boemer}      % badge no. 0043286
\author[3]{Ryan~N.~Coffee}          % badge no. 0035049
\author[3,4]{Nick~Hartmann}         % badge no. 0041641
\author[5]{Rupert~Heider}           % badge no. 0044480
\author[5]{Martin~S.~Wagner}        % badge no. 0044484 % affiliation to be checked
\author[5,6,b]{Wolfram~Helml}       % badge no. 0044485 % TUM, LMU, now at DELTA, TU Dortmund
\author[7,8]{Tetsuo~Katayama}       % badge no. 0018149
\author[9,c]{Tokushi~Sato}          % badge no. 0031381 % was at CFEL, now at European XFEL
\author[3]{Takahiro~Sato}           % badge no. 0011412 % LCLS?
\author[7,8]{Makina~Yabashi}        % badge no. 0044484 % affiliation to be checked
\author[1,2,*]{Christian~Bressler}  % badge no. 0031361 
\affil[1]{European XFEL GmbH, Holzkoppel 4, 22869 Schenefeld, Germany}
\affil[2]{The Hamburg Centre for Ultrafast Imaging, Luruper Chaussee 149, 22761 Hamburg, Germany}
\affil[3]{SLAC National Accelerator Laboratory, 2575 Sand Hill Rd., Menlo Park, CA 94025, USA}
\affil[4]{Coherent, Inc., 5100 Patrick Henry Dr., Santa Clara, CA 95054, USA} 
\affil[5]{Physik-Department E11, Technical University of Munich, James-Frank-Str. 1, 85748 Garching, Germany}
\affil[6]{Faculty of Physics, Ludwig-Maximilians-Universität Munich, Am Coulombwall 1, 85748 Garching, Germany} % ? full address
\affil[7]{Japan Synchrotron Radiation Research Institute, 1-1-1 Kouto, Sayo-cho, Sayo-gun, Hyogo 679-5198, Japan}
\affil[8]{RIKEN SPring-8 Center, 1-1-1 Kouto, Sayo-cho, Sayo-gun, Hyogo 679-5148, Japan}
\affil[9]{Center for Free-Electron Laser Science, Deutsches Elektronen-Synchrotron DESY, Notkestraße 85, 22607 Hamburg, Germany}
\affil[a]{now at Deutsches Elektronen-Synchrotron DESY, Notkestraße 85, 22607 Hamburg, Germany}
\affil[b]{now at Center for Synchrotron Radiation, Technical University of Dortmund, Maria-Goeppert-Mayer-Str. 2, 44227 Dortmund, Germany}
\affil[c]{now at European XFEL GmbH, Holzkoppel 4, 22869 Schenefeld, Germany}
\affil[*]{corresponding authors: michael.diez@xfel.eu, sebastian.schulz@desy.de, christian.bressler@xfel.eu}


\begin{document}
\begin{abstract}
%\section*{ABSTRACT}
\input{main-abstract.tex}
\end{abstract}
%%  create title and ensure proper formatting of the whole page
%
\flushbottom
\maketitle
\thispagestyle{empty}

%%  introduction
%
\section*{Introduction}
%
%   include the introduction from an external file
\input{main-introduction.tex}

%%  experimental details - key aspect of the manuscript -> separate section
%
\section*{Experimental details}
%
%   include the experimental details from an external file
\input{main-experiment.tex}

%%  results
%
\section*{Results and Discussion}
%
%   include the results and their discussion from an external file
\input{main-results.tex}

%%  discussion
%
\section*{Conclusions}
%
%   include the conclusions from an external file
\input{main-conclusions.tex}

%%  methods
%
\section*{Methods}
%
%   include the methods from an external file
\input{main-methods.tex}
\section*{Data and code availability}
The data and code that supports the findings of this study are available from the corresponding author(s) upon reasonable request.

%%  references / bibliography
%
% \nocite{*}
\bibliography{main-bibliography}

%%  acknowledgements
%
\section*{Acknowledgements}
This work is supported by the Deutsche Forschungsgemeinschaft (DFG) via the Cluster of Excellence `The Hamburg Centre for Ultrafast Imaging' -EXC1074- project ID~194651731 and via SFB925 (TP A4), and by European XFEL.
The flat sheet liquid jet system was financed by the German Federal Ministry of Education and Research (BMBF, contract no.~05K16PE1).
S.S., A.G., and C.Br. acknowledge support from the European Cluster of Advanced Laser Light Sources (EUCALL) project which had received funding from the European Union's Horizon 2020 research and innovation programme under grant agreement No~654220.
The XFEL experiments were performed at the BL3 of SACLA with the approval of the Japan Synchrotron Radiation Research Institute (JASRI) (Proposal No. 2017A8072). The authors thank Prof. Thomas Feurer for fruitful discussions, whose insights into optics and related experimental fields made this experiment possible.

%%  author contributions
%
\section*{Author contributions statement}
M.D., S.S., and A.G. planned and coordinated the measurement campaign.
C.Br. and A.G. conceived the in-situ experiment with sample and timing tool.
N.H., A.G. and R.N.C. conceived the interferometric setup as a sensitive timing tool.
S.S. prepared and characterised the flat-sheet liquid jet system.
M.D., A.G., S.S., C.Bö., N.H., W.H., R.H. and M.S.W. performed the experiments at the SACLA XFEL facility, including the setup of the optical timing tool instrumentation.
T.K., To.S.,  Ta.S. and M.Y. were involved in beamline control and integration.
M.D. analysed the data, M.D., S.S., A.G. and C.Br. interpreted the data, and M.D., S.S. and C.Br. wrote the manuscript with contributions from all authors.

%%  additional information
%
\section*{Additional information}

%   competing interests statement
%   note from the original template: The corresponding author is responsible for submitting a competing interests statement on behalf of all authors of the paper. This statement must be included in the submitted article file.
%   additional information here: https://www.nature.com/nature-research/editorial-policies/competing-interests
%
\subsection*{Competing interests}
The author(s) declare no competing interests.

%%  figures, combined at the end of the manuscript for initial submission
%
\input{main-figures.tex}

\end{document}

%% file: main-abstract.tex
% !TeX root = ./main.tex
%
%%  abstract

We present a novel, highly versatile, and self-referenced arrival time monitor for measuring the femtosecond time delay between a hard X-ray pulse from a free-electron laser and an optical laser pulse, measured directly on the same sample used for pump-probe experiments.
Two chirped and picosecond long optical supercontinuum pulses traverse the sample with a mutually fixed time delay of 970~fs, while a femtosecond X-ray pulse arrives at an instant in between both pulses.
Behind the sample the supercontinuum pulses are temporally overlapped to yield near-perfect destructive interference in the absence of the X-ray pulse.
Stimulation of the sample with an X-ray pulse delivers non-zero contributions at certain optical wavelengths, which serve as a measure of the relative arrival time of the X-ray pulse with an accuracy of better than 25~fs.
We find an excellent agreement of our monitor with the existing timing diagnostics at the SACLA XFEL with a Pearson correlation value of 0.98.
We demonstrate a high sensitivity to measure X-ray pulses with pulse energies as low as 30~$\upmu$J.
Using a free-flowing liquid jet as interaction sample ensures the full replacement of the sample volume for each X-ray/optical event, thus enabling its utility even at MHz repetition rate XFEL sources.
%
%% end of file

%% file: main-introduction.tex
% !TeX root = ./main.tex
%
%%  introduction

Since the advent of hard X-ray free-electron lasers (XFEL) about ten years ago\cite{Emma:NPhoton:2010:LCLSFirstLasing,Ishikawa:NPhoton:2012:SACLAFirstLasing,Kang:NPhoton:2017:PALXFELTimingJitter,Milne:ApplSci:2017:SwissFELProject,Decking:NPhot:2020:XFELCommissioning}, a scientific community has grown seeking to exploit their intense femtosecond pulses to investigate ultrafast dynamics in optical pump/X-ray probe experiments\cite{Zhang:Nature:2014:ChargeSpinDynamics,Lemke:NComms:2017:SpinStateSwitching,Chapman:Nature:2011:ProteinCrystallography}.
However, despite having ultrashort femtosecond X-ray and optical pulses available, the level of synchronisation between the two independent light sources limits the achievable time resolution in these experiments.
Timing jitter at large-scale XFEL facilities arises from different sources, most notably in the linear accelerator itself\cite{Lohl:PRL:2010:SyncBAMProof}, the reference signal distribution scheme across the facility\cite{Xin:Optica:2018:SyncSystems}, and the optical laser, including length variations of the optical beam paths\cite{Cinquegrana:PhysRevSTAB2014:OpticalBeamTransport}.
In order to minimise the timing jitter between optical lasers and the X-ray pulses at the experimental stations, but also to stabilise the linear accelerator, different advanced synchronisation schemes have been developed and implemented\cite{Lohl:PRL:2010:SyncBAMProof,Byrd:IPAC:2010:LCLSLaserSync,Schulz:NComms:2015:FLASHOverallSynchronization,Viti:ICALEPCS:2018:BAM,Xin:Optica:2018:SyncSystems}, for instance utilising transit time-stabilised optical fibres to link the photo-injector laser of the electron gun, the reference signals of the acceleration modules and the remote optical laser system used for sample excitation to a common optical reference.
While the synchronisation accuracy was pushed to a level of about ten femtoseconds between the facility's reference oscillator and the subsystem oscillators, such as the seeder of the pump-probe laser system, the overall level of synchronisation was found to be on the same order of magnitude as the few ten femtosecond pulse duration at a soft X-ray FEL\cite{Schulz:NComms:2015:FLASHOverallSynchronization}.\newline
For time-resolved experiments utilising optical pump and X-ray probe pulses, usually optical pulse energies of several $\upmu$J up to mJ are necessary for sample excitation\cite{Bressler:ChemRev:2004:XAS}, which consequently require the use of amplified laser systems.
However, the design of the laser amplifier and compressor can introduce a few femtoseconds to up to hundreds of femtoseconds additional timing jitter\cite{Prinz:OptExpr:2014:OPCPAJitter, Casanova:OptLett:2016:AmplifierJitter,Klingebiel:OptExpr:2012:CompressorJitter}.
Subsequent optics for tailoring and guiding the optical laser beam to the sample interaction region introduce further timing jitter between the optical laser and the X-ray pulses.\newline
Ultimately, additional diagnostic methods are therefore used to measure the relative arrival time between the optical and X-ray pulses in order to minimise arrival time uncertainties, which cannot be removed.
These so-called timing tools are meanwhile implemented at most XFEL beamlines and instruments\cite{Bionta:OptExpr:2011:SpectralEncodingLCLS,Schorb:ApplPhysLett:2012:LCLSCrossCorrelatorGasPhase,Krupin:OptExpr:2012:LCLSCrossCorrelator,Hartmann:NPhoton:2014:LCLSSpectrogramTiming} and allow determining the arrival time for each laser pump/X-ray probe event. As such they provide means to sort the data \textit{a posteriori}, yielding a superior timing precision in the analysis of actual pump-probe experiments following this correction\cite{Harmand:NPhoton:2013:LCLSTimeSortingBismuth,Hartmann:NPhoton:2014:LCLSSpectrogramTiming}.
Different schemes are employed for an accurate arrival time measurement including phase-dependent streaking of photo-ionised atoms with an overlapping THz electric field\cite{Fruhling:NPhoton:2009:FLASHTHzStreaking,Grguras:NPhoton:2012:LCLSTHzStreaking,Gorgisyan:OptExpr:2017:SwissFELTHzStreaking}, but often also exploiting the transient X-ray-induced refractive index changes, which affect both the transmission and reflection of an optical laser pulse through the chosen material.
Two straightforward techniques have been implemented at different XFEL facilities, known as spatial encoding and spectral encoding.
In both methods the X-ray pulse induces a change of the optical properties of a sample, such that the resulting change of transmission and reflectivity of an optical light pulse is used to determine the relative arrival time of the two pulses.
Spatial encoding retrieves the arrival time information from the lateral modulation of a reflected or transmitted optical beam profile, when the X-ray beam crosses with the optical beam at a given angle on the interaction sample\cite{Beye:ApplPhysLett:2012:LCLSSpatialEncoding,Riedel:NatComms:2013:FLASHSpatialEncoding,Sato:ApplPhysExpr:2014:SACLATimingTool}.
Spectral encoding uses a chirped laser pulse, where frequencies ranging from lower to higher optical frequencies---or from red to blue in spectral colour---arrive at well-defined and distinctly different times.
The transient refractive index in the material induced by the X-ray pulse therefore encodes the arrival time as a spectral feature, for example a step-like change of the transmitted laser spectrum at a specific wavelength\cite{Bionta:OptExpr:2011:SpectralEncodingLCLS,Bionta:RevSciInstr:2014:LCLSSpectralEncoding,Coffee:RSTA:2019:UltrafastCapabilitiesLCLS}.
\newline
In this work, we present a variation of the spectral encoding approach, which aims to reliably operate at all XFEL facilities, in particular at high-repetition rates and at the same time accepting a broad range of X-ray pulse intensities.
We modified this scheme to better exploit the small changes of the refractive index of the material induced by the transmitting X-ray pulse, which enhances the sensitivity towards highly transparent materials including diamond foils and few micrometer thin flat sheet liquid jets.
For this purpose we add an interferometric detection scheme, which is not only sensitive to the potentially weak X-ray induced absorption change, but also to the X-ray-induced phase change of the transmitted chirped optical pulse.
This strategy opens up a larger choice of materials for precise timing tool measurements, including those which are not feasible in either spatial or spectral encoding schemes like diamond.
The major improvement is a direct result of its background-free detection principle, due to near-complete annihilation of the entire signal by destructive interference in absence of an X-ray pulse impinging on the sample.
This condition can be fulfilled for nearly all transparent samples, and easily with optically isotropic materials.
The presence of a femtosecond X-ray pulse on the material yields a non-zero interferometric signal, which can be analysed to extract the precise relative timing between individual X-ray and optical pulse pairs.
As we demonstrate in this work for individual chirped pulses, this scheme operates reliably, even if the material would vary its optical properties from shot to shot.
This self-referencing method always yields complete destructive interference when the optical properties of the interfering beams are identical, regardless of when they would gradually change from pulse to pulse. 
\newline
To benchmark the capabilities of the interferometric approach we carried out a proof-of-principle experiment at the SPring-8 Angstrom Compact free-electron LAser (SACLA) in Japan, where we used a free-flowing flat sheet liquid jet\cite{Ekimova:StructDyn:2015:FlatSheetJet} as the interaction sample.
This approach has the additional benefit of being applicable to next generation MHz repetition rate XFEL sources with few-mJ pulse energies, delivering up to several kilowatts of X-ray pulse power\cite{Altarelli:NIMB:2011:XFELProject,Decking:NPhot:2020:XFELCommissioning,Halavanau:JSR:2019:LCLSIIHighPower}, where the established timing tools can suffer from these high repetition rates.
The main concern are pile-up effects caused by the mJ-level XFEL pulses on the timing tool materials, which deteriorate the extracted signal during the MHz repetition rate pulse delivery due to saturation effects caused by the previous pulses, either by the accumulated heat load or by excited state carrier lifetimes exceeding the time interval between pulses.
Furthermore, commonly used samples such as Si${}_3$N${}_4$ also will be severely damaged after only a few seconds exposure, even at low repetition rates\cite{Robinson:JPhysB:2015:SACLAParticleImaging}.
Consequently, for fully exploiting the MHz capabilities at the latest and upcoming XFEL sources, samples are required which are capable to distribute the X-ray induced heat load fast enough to guarantee no heat pile-up between the pulses.
Alternatively the sample volume needs to be replaced with a fresh sample volume between two pulses.
By using a fast-flowing liquid jet we are able to replace the entire sample volume for each X-ray pulse even at MHz repetition rates, thus evading the problem of destroying or deteriorating the interaction sample. 
A combination of a fast-flowing liquid jet, which can reach speeds of around 100~m/s (refs.\cite{Galler:JSR:2019:FXE,Khakhulin:ApplSci:2020:FXE}), and sufficiently small optical laser and X-ray foci of well below 50~$\upmu$m, allow the exchange of the sample volume at MHz repetition rates.
Crucially, we also demonstrate the capability to apply this interferometric encoding approach as an in-situ timing tool on the sample under investigation.
For this, we superimpose an additional intense optical laser pulse with a wavelength of 400~nm, which serves as a pump pulse in an actual liquid chemistry experiment. 
The in-situ arrival time measurement could potentially increase the achievable time resolution by also measuring additional timing jitter occurring at the sample position, e.g. spatial displacement of the sample due to nearby vibrational sources or using a liquid jet.
%
%%  end of file

%% file: main-experiment.tex
% !TeX root = ./main.tex
%
%%  experimental details
%
The experiments were performed in the EH2 experimental hutch of SACLA.
This XFEL delivers femtosecond X-ray pulses at a repetition rate of either 30~Hz or 60~Hz.
The photon energy can be adjusted in the 4~keV to 10~keV range with up to 0.6~mJ pulse energy and, with decreasing pulse energy, up to 20~keV can be reached\cite{Yabashi:ApplSci:2017:SACLAStatus}.
Throughout the experiment, the repetition rate was set to 30~Hz and the photon energy to 5.2~keV, with a pulse energy of about 300~$\upmu$J.
The X-ray beam spot at the sample position was set to a circular radius of 100~$\upmu$m, using the facility's compound refractive lenses \cite{Katayama:JSR:2019:SaclaCLR}.\newline
The facility's optical laser system is based on a commercial Ti:sapphire chirped-pulse amplification (CPA) system operating at 1~kHz (Micra and Legend Elite, Coherent Inc.), which delivers pulses with a duration of 25~fs FWHM and 15~mJ of energy at a central wavelength of 800~nm \cite{Tono:NJP:2013:SACLABeamlines,Yabashi:JSR:2015:SACLAFacility}.
It is synchronised to the accelerator's master clock by locking the oscillator cavity length using a commercial standard phase-locked loop\cite{Tono:NJP:2013:SACLABeamlines} (Synchrolock-AP, Coherent Inc.).
A pulse picker selects a subset of optical pulses from the 1~kHz laser amplifier output to match the X-ray repetition rate.
The time delay between the optical laser and X-ray pulses can be freely adjusted with sub-picosecond accuracy using a trigger and clock delay module (84DgR5C01, CANDOX Systems Inc.), while for femtosecond timing an additional optical delay line is used.
For our experiment, schematically shown in Fig.~\ref{fig:interferometric_setup}, we split the optical laser beam into three paths.
The first one is used for the operation of the facility's standard timing tool (TT), which is based on the spatial encoding technique and located a few metres upstream of the experimental hutch EH2\cite{Sato:ApplPhysExpr:2014:SACLATimingTool,Katayama:StructDyn:2016:SACLATimingToolBranch}.
The second beam is used for exciting the actual sample in our pump-probe geometry and for the calibration of our interferometric timing tool.
For this, the 800~nm optical pulse is frequency-doubled (second-harmonic generation, SHG) in a BBO crystal to a wavelength of 400~nm.
Finally, the third beam is used for the simultaneous measurement of the X-ray/optical relative arrival time at the sample position and on the sample itself using our novel interferometric technique.
For this, it is further split into two beams in a common path interferometer, which had already been proposed in 1958 by Mertz\cite{Mertz:JPRadium:1958:Spectrometer} and was first applied in astronomy.\cite{AHearn:ApplOpt:1974:Spectrometer}
In the interferometer, both laser pulses travel along the same optical path and traverse the same optical elements.
Therefore, the extracted interferometric signal is insensitive to environmental disturbances, such as air flow, temperature variations and vibrations.
For the interferometric scheme, 800~$\upmu$J of the 800~nm fundamental laser pulses are focused into a 1.5~m long tube filled with 1.5~bar pressurised Argon gas, generating a broadband supercontinuum pulse\cite{Trushin:OptLett:2007:Filamentation} ranging from 300~nm to over 1000~nm in wavelength.
The Argon cell and several transmissive optics in the beam path exhibit sufficient dispersion to chirp the continuum to a pulse duration of approximately 2.5~ps within a narrow wavelength range of about 350-430~nm.
The supercontinuum laser pulse is guided over an optical delay line ($\Delta t_\text{SC}$) to the interferometer.
To prevent linear or non-linear sample excitation by the supercontinuum pulse, a BG38 color glass filter is used to block the 800~nm contribution from the supercontinuum, reducing the supercontinuum pulse energy to below 1~$\upmu$J.\newline
Its polarisation is set to 45${}^\circ$ via the broadband polariser P$_1$ and it is subsequently guided through a 4~mm thick a-cut alpha barium borate crystal ($\alpha$-BBO${}_1$).
The optical axis of the crystal is set to be horizontal, which evenly projects the incoming pulse into a vertical and horizontal polarisation component, due to the birefringence of the barium borate crystal.
By this, the extraordinary vertical polarised pulse exits the crystal about 970~fs ahead of the ordinary horizontal polarised pulse (Fig.~\ref{fig:interferometric_setup}~a).
This pair of orthogonal polarised supercontinuum pulses is focused to a 60~$\upmu$m circular spot onto the sample, spatially overlapping with the 100~$\upmu$m circular X-ray pulse.
The arrival of the X-ray pulse is temporally set in between both supercontinuum pulses by adjusting the delay line in the supercontinuum beam path ($\Delta t_\text{SC}$), and marks time zero by stimulating the sample (Fig.~\ref{fig:interferometric_setup}~b).
%By absorbing X-ray photons, high-energy electrons in the conduction band of the sample are generated.
%Within tens of femtoseconds, these high-energy electrons relax to the bottom of the conduction band, creating secondary charges in an electron cascade\cite{Medvedev:ApplPhysB:2015:ElectronKinectics}.
By absorbing X-ray photons, high-energy electrons in the conduction band of the sample (either a reference solid, or the liquid water jet) are generated. Within tens of femtoseconds, these high-energy electrons create secondary charges in an electron cascade \cite{Medvedev:ApplPhysB:2015:ElectronKinectics}. In solids (e.g., the SiO$_2$ plate used for calibration) these electrons relax energetically to the bottom of the conduction band, in liquid water these electrons thermalize within 300~fs as solvated electrons \cite{LaVerne:JPhysChemA:2005:HeavyIonRadiolysis}.
%LaVerne:JPhysChemA:2005:HeavyIon Radiolysis: J. A. LaVerne, I. Stefanic, S. M. Pimblott, J. Phys. Chem. A 109, 9393-9401 (2005)
This process changes the band structure of the material, thus changing its refractive index.
The spectral regions of each supercontinuum pulse, which spatially overlap in the sample with the X-ray pulse and temporally trail behind the exciting femtosecond X-ray pulse experience an amplitude reduction due to the altered absorption cross section and a phase modulation due to the change in refractive index.
Both supercontinuum pulses are then transmitted through a second, identically a-cut barium borate crystal ($\alpha$-BBO${}_2$) with its optical axis, however, set to be vertical.
As a result, both supercontinuum pulses overlap temporally again, but maintain their previously generated orthogonal polarisation (Fig.~\ref{fig:interferometric_setup}~c).
By precisely adjusting the rotation of the second $\alpha$-BBO${}_2$ crystal around the vertical optical axis, the effective length of the crystal can be changed, which allows to exactly establish a spectral offset of $\lambda/2$ between both pulses.
Utilising a polariser set to 45${}^\circ$ with respect to the two axes of polarisation, half of the amplitude of each pulse is projected onto the same polarisation plane. 
Due to destructive interference between both supercontinuum pulses, nearly complete annihilation over all contributing wavelengths can be achieved in the absence of the X-ray pulse.
Crucially, when X-rays interact with the sample, a residual non-zero interferometric contribution is observed in those central wavelength regions, where only one of the two supercontinuum pulses experienced the modified refractive index (coloured portion of the pulse in Fig.~\ref{fig:interferometric_setup}~c).
The spectral wings of each pulse experience the same sample condition, either with the unperturbed refractive index on the red side of the spectrum or the X-ray-altered refractive index on the blue side of the spectrum, thus completely annihilate (indicated by the grey wings in Fig.~\ref{fig:interferometric_setup}~c).
The remaining central portion in the presence of an X-ray pulse does not interfere destructively and results in the indicated residual spectrum (Fig.~\ref{fig:interferometric_setup}~d).
This X-ray-induced remaining interferometric signal is guided to a spectrograph (Andor Shamrock 193i, focal length 193~mm, 600~grooves/mm grating with a blaze wavelength of 500~nm), where the dispersed optical spectrum is recorded with a pixel detector (Adimec Opal, 1920$\times$1080~pixels, pixel size 5.5~$\upmu$m${}^2$).
The high dispersion of the chosen grating limits the recorded spectral region to about $\Delta\lambda\,\approx\,$100~nm around 400~nm, and defines ultimately the temporal window of the measurement, as well as its final time resolution.\newline
In an actual scientific experiment, where the interferometric timing tool is used as an in-situ arrival time monitor, the intense 400~nm pulse is used to excite the sample. Thus, an additional optical pulse arrives together with the supercontinuum double pulses and the X-ray pulse in a narrow time window. This scenario is shown in Fig. \ref{fig:interferometric_setup}~e).\newline
The calibration of the spectrometer was accomplished with a calibrated Hg/Ne light source.\newline
As a sample delivery system we used a commercial free-flowing liquid flat sheet jet system (see Methods and ref.\cite{Ekimova:StructDyn:2015:FlatSheetJet}) to provide a fresh sample volume for each X-ray/optical pulse event.
Compared to commonly used round liquid jets, a flat sheet surface yields a more predictable beam path of the optical light through the sample for all involved optical beams.
%
%%  end of file

%% file: main-results.tex
% !TeX root = ./main.tex
%
%%  Results section

\subsection*{Time calibration and data analysis}
Upon sample excitation, the interferometric signal spans over a defined range of wavelengths, limited by two sharp boundaries, created by the interaction of the X-ray pulse with the sample (see schematic in Fig.~\ref{fig:interferometric_setup} d).
The width of the spectrum corresponds to the fixed time delay between both supercontinuum pulses governed by the thickness and rotation of the birefringent barium borate crystals, which in our case is 970~fs.
Both the red and blue cut-off wavelengths deliver the same information about the relative arrival time of the X-ray pulse with respect to the supercontinuum pulse pair.
This redundancy can be used to enhance the timing accuracy, as will be shown below.\newline
In a timing jitter measurement, one records for each X-ray pulse an optical spectrum with a constant width, but whose central portion moves together with its boundaries to lower and higher wavelengths, depending on the temporal overlap of each individual X-ray pulse with the supercontinuum pulse pair in the sample.
The arrival time for each X-ray pulse can be retrieved by precisely determining the spectral wavelength position of each rising and falling edge of the interferometric signal.
For this, it is required to map each optical wavelength, and, thus, each detector pixel of the spectrometer into the time domain.
The resulting time axis is a function of the chirp of the supercontinuum pulse.
To reliably convert the wavelength, or detector pixel values, to a time axis, we operated the interferometric setup with the optical 400~nm excitation laser pulse instead of the X-ray pulse.
Since the supercontinuum and the 400~nm pulses originate from the same source, their residual timing jitter is sufficiently small, such that the arrival time between these pulses can be precisely adjusted with an optical delay line ($\Delta t_2$ in Fig.~\ref{fig:interferometric_setup}).
We recorded 50 individual interference spectra for each of the 48 different time delays in 100~fs steps, corresponding to 15~$\upmu$m steps of the delay line.
Four time delays are illustrated in Fig.~\ref{fig:time_calibration_2dimg} a).
Polycrystalline SiO${}_2$ with a thickness of 90~$\upmu$m was used for these calibration measurements instead of the flat sheet jet.
The solid sample was chosen for its simple setup, but also to eliminate additional timing jitter, which potentially could be caused by a fluctuating liquid jet. 
In a stark contrast to the X-ray induced interferometric timing signal illustrated in Fig \ref{fig:interferometric_setup} d), the laser induced interferometric timing signal looks different.
Due to the large bandgap of SiO$_2$ of 8.9~eV, a non-linear absorption of the 400~nm optical pump pulse is needed to generate one conduction band electron. 
Thus, the optical induced electron density is smaller than the X-ray induced electron density, resulting in a weaker, nearly not detectable, interferometric timing signal. 
The optical signal is dominated by two coherent artifacts at the spectral position where the leading and trailing edges of the X-ray induced interferometric signal would be expected.\newline
The mapping of wavelengths (or detector pixel) into the time domain is determined by evaluating the edge position for each spectrum of a single time delay, using Eq.~(\ref{eqn:fit}) to fit the edges, as shown in Fig.~\ref{fig:time_calibration_2dimg}~b).
The analysis of all 50 shots at one fixed time delay yields an error of $\Delta$cen$_l$=2.2~pixels FWHM for the leading edges and $\Delta$cen$_t$=1.5~pixels FWHM for the trailing edges (indicated by the vertical shaded rectangles in the example in Fig.~\ref{fig:time_calibration_2dimg}~c).
Together with the error bars of the error function center parameter, the overall uncertainty of the time calibration accumulates to $\pm 6$~fs FWHM. 
The entire set of calibration points for converting wavelengths into a relative time delay for each leading and trailing edge is fitted with a second order polynomial function, where time zero is set arbitrarily to the centre of the detector. It delivers the calibration curve for this timing tool (Fig.~\ref{fig:time_calibration_2dimg}~d), which results in the aforementioned wavelength-dependent accuracy to determine a time point within a precision of $\pm$2~fs for any given detector pixel.\newline
%As will be shown in the next section, the interferometric timing signal produced by the optial laser in the time calibration looks different to the 

\subsection*{Arrival time measurements}
An actual liquid chemistry experiment was prepared by using a 100~mM aqueous solution of sodium iodide as the target sample, which upon focused 400~nm optical laser excitation generates nascent iodine radicals\cite{Pham:JAmChemSoc:2011:Iodide}.
The goal of this experiment is to demonstrate the reliable operation of the interferometric timing scheme in the same sample used for the scientific experiment, in particular that it delivers accurate arrival time information for each X-ray pulse in the presence of an additional intense laser stimulus.
Therefore, this scheme will provide additional flexibility towards a variety of samples used in liquid phase structural dynamics experiments.
Since we expected an increased sensitivity of this scheme to extract accurate arrival time information, we attenuated the X-ray pulses with a 50~$\upmu$m thick Al foil, transmitting only one tenth of the original pulse energy, i.e. 30~$\upmu$J.
Even under these conditions and using a relatively large X-ray spot size of 100~$\upmu$m on the sample and a corresponding X-ray power density of 2~TW/cm${}^2$, a clear timing signal is measured, shown in Fig.~\ref{fig:run550234_raw}~a).
The simultaneously recorded data of the established SACLA timing tool based on spatial encoding  (Fig.~\ref{fig:run550234_raw}~b) has been analysed using Eq.~(\ref{eqn:fit}) without the sine contribution (see Methods and ref.\cite{Sato:ApplPhysExpr:2014:SACLATimingTool}) to determine the edge positions and thus the relative arrival times.
With the reported\cite{Katayama:StructDyn:2016:SACLATimingToolBranch} dimensions of the line focus for the SACLA spatial encoding scheme (3~$\upmu$m~$\times$~780~$\upmu$m), using roughly two percent of the original X-ray pulse energy split from the main beam with a silicon grating, the X-ray power density of 5~TW/cm${}^2$ is higher than the 2~TW/cm used for the interferometric timing tool.
In this comparison study, 5000 individual X-ray pulses were recorded with each timing tool.
In Fig.~\ref{fig:run550234_raw}~c) five consecutive time arrival measurements with the interferometric timing tool are shown.
The identical X-ray pulse arrival time measurements with the SACLA timing tool is shown in Fig.~\ref{fig:run550234_raw}~d).\newline
We observe a very good correlation (Fig.~\ref{fig:run550234_ana}) with a Pearson correlation value\cite{Spiegel:book:2017:Statistics,Freedman:book:2007:Statistics} (see Eq.~(\ref{eqn:pearsoncorrelation}) in Methods) of 0.98 for 89\% of the data, when comparing the arrival times measured with both timing tools.
The remaining 11\% of the data did not yield a trustworthy measurement of the timing signal, mainly due to occasional low X-ray pulse energy shots for the spatial encoding scheme, but also due to occasional and random disturbances of the flat sheet liquid jet, which scattered the optical beam away from the spectrometer entrance. 
Those X-ray arrival time measurements where the parameter defining the center position of the fitting function, in either of the two timing tools, had an error larger than 3$\sigma$, where rejected from the analysis.\newline
In a perfect correlation, i.e. a Pearson value of 1, all data points would lie on the orange diagonal line in Fig.~\ref{fig:run550234_ana}, such that any deviation from this line is a measure for the uncertainty between the two schemes, resulting to about 39~fs FWHM (inset in Fig.~\ref{fig:run550234_ana}).
%This residual timing jitter can be dominantly attributed to mechanical vibrations and air-induced fluctuations in the several metres long different optical beam paths from the laser system to the two measurement positions. 
A major contribution to the residual timing jitter can be caused by the flat sheet liquid jet. In such systems with connections to powerful liquid pumps, mechanical vibrations can never be fully prevented. Spatial oscillations of the liquid flat sheet along the X-ray axis can introduce an additional timing jitter, not recognised by the remote SACLA timing tool. In our configuration, with an angle between the X-ray and optical pulses of 10$^\circ$, a spatial oscillation of $\pm350\,\upmu$m would be enough to introduce the measured residual timing jitter. \newline
The arrival time distribution itself (or the jitter) measured by the interferometric encoding and SACLA spatial encoding deliver the histograms shown in Fig.~\ref{fig:run550234_ana} with almost identical FWHM values of (513~$\pm$~22)~fs and (510~$\pm$~39)~fs for the interferometric encoding and the spatial encoding scheme, respectively.
The relative large error in this particular measurement is caused by the low X-ray pulse energy which in turn generates timing signals with a rather small amplitude, thus the fitting function residuals are larger.
In addition, we confirmed that without the Al attenuator the interferometric signal strength increases by the same factor of ten as the X-ray pulse intensity, thus confirming operation in the linear regime.\newline
We believe that the interferometric encoding measurement should deliver a more precise arrival time information for a pump-probe experiment, since it operates directly on the sample.
This could be demonstrated in an experiment, where the selected sample develops a very steep femtosecond response upon photo-excitation.
Correcting the measured time trace for every timing point from each timing tool would then deliver a signal with possibly different rise times.
The timing tool yielding the shorter rise time would then quantify its increased accuracy.
Although such a study could not be performed in the current experimental campaign, we first confirmed its utility for this situation as stand-alone timing tool at the sample position.
We guided an intense ($\approx$100~TW/cm${}^2$) 400~nm optical beam onto the liquid jet together with the X-ray pulses in order measure the relative arrival time between the 400~nm pump and X-ray probe beams.
In this arrangement, the X-ray pulse, which is used as a pump pulse for the arrival time measurement, simultaneously would serve as the probe pulse during a liquid chemistry experiment, where the dynamics in the sample are initiated with the additional intense 400~nm pulse.
The X-ray arrival time monitor then needs to be operational even in the presence of that additional intense 400~nm pump pulse, which could potentially distort the arrival time measurement.
To investigate the influence of the additional 400~nm stimulus, we performed another arrival time measurement with fixed timing between the X-ray and supercontinuum double pulses, while scanning the additional intense pump pulse from $-$1,500~fs to 1,000~fs (Fig.~\ref{fig:run550236_full}~a).
For each time delay setting, 200~supercontinuum probe pulses were recorded.
The X-ray induced arrival time regions are framed in orange (leading edges) and purple (trailing edges).
The intense 400~nm pulses act similar to the X-ray pulses, such that each pulse creates its own distinct plateau shape in the supercontinuum double pulse interferometric timing tool due to the altered transient refractive index of the sample.
The leading (pink) and trailing (gray-blue) edges of the 400~nm stimulus timing signal are indicated in a) for each single pulse.
Since the 400~nm and supercontinuum pulses used for the timing tool are generated from the same source, they are practically jitter free, and, thus the recorded timing jitter of the 400~nm induced timing signal should be very small.
In consequence, both timing signals can overlap during a time delay scan, which is most apparent when their respective edges cross each other, i.e. when both stimulating pulses (400~nm pump and X-ray) arrive simultaneously.
In Fig.~\ref{fig:run550236_full}~a), three distinct delay times for the 400~nm pump beam are indicated with white boxes.
At each of these delay settings, either the trailing edge, both edges or the leading edge of the X-ray-induced timing signal overlaps with one or both edges of the optically induced timing signal. 
This initially leads to an increased uncertainty in the determination of those X-ray arrival times (Fig.~\ref{fig:run550236_full}, right panels).
The distributions in the right panel show, similar to the inset in Fig.~\ref{fig:run550234_ana}, the residual jitter between the interferometric timing tool and the SACLA timing tool, which again serves as reference.
We found the accuracy limited to about twice the individual uncertainty of approximately 20~fs, however, further analysis can mitigate this increased uncertainty, either by simply selecting the unaffected edge to analyse or by adding more detailed knowledge into the edge shape do distinguish the 400~nm stimulus timing signal from the X-ray timing signal, when both are overlapped.
The latter is only absolutely required at the point in time, where the X-ray and the 400~nm pulses arrive simultaneously.
In all other cases where the X-ray pulse and 400~nm stimulus pulse are separated by more than $\approx$20~fs, we always find an undisturbed edge for a precise arrival time determination, as shown in Fig.~\ref{fig:run550236_full}~d) and~e).
We analysed the arrival times of the 400~nm optical pulses for each delay in a similar way we analysed the X-ray pulses, with the resulting edge positions indicated by the dots in Fig.~\ref{fig:run550236_full}~a).
The edge position temporal jitter is shown in Fig~\ref{fig:run550236_full}~h) for all leading (pink) and trailing (gray-blue) edges, within the single delays.
When the 400~nm stimulus is set to arrive approximately 1~ps earlier or later in time compared to the X-ray pulses, the arrival time jitter of the 400~nm pulses is 7~fs FWHM and by this similar to the calibration error without X-ray pulses.
As the relative time delays between X-ray and 400~nm stimulus approach each other it becomes more and more challenging to disentangle the respective timing signals, which leads to an increased uncertainty in measuring the X-ray pulse arrival times as described before, but also in measuring the 400~nm stimulus arrival signals.
The maximal uncertainty, when both pulses are exactly overlapped in time, is approximately 45~fs FWHM.
A double Gaussian function was fitted to the data in Fig.~\ref{fig:run550236_full}~h) with the two peaks centered at the mean edge positions of the leading and trailing interferometric X-ray induced timing edges.
The widths of the two Gaussians are fixed to the experimentally measured FWHM arrival time jitter of the X-ray pulses of 513~fs.
In addition, one can even use the 400~nm stimulus timing signal imprinted in the X-ray arrival time measurement to obtain the actual detector pixel, i.e. wavelength, to time mapping.
This could potentially be useful in certain measurement campaigns, where data is accumulated over a long time and delay scans are repeated to gain more statistics.
In such a case a change in the chirp of the supercontinuum could potentially be recognized during the measurement and the time calibration could be adjusted accordingly without the need to abort the actual experiment.\newline
Another notable feature of the recorded interference pattern of the supercontinuum double pulse is that it contains in-situ information about the actual flat sheet thickness of the free-flowing liquid jet.
This information is obtained within the picosecond short time span of the supercontinuum passing through the flat sheet jet. 
% original:
% Close to the leading and trailing edges additional oscillations caused by thin film interference are observed (e.g. Fig.~\ref{fig:time_calibration_2dimg}~b,c), which we use to extract the liquid jet thickness (see Eq.~(\ref{eqn:thicknessretrieval}) in Methods).
% new:
Close to the leading and trailing edges additional oscillations caused by thin film interference are observed (e.g. Fig.~\ref{fig:time_calibration_2dimg}~b,c), which we use to extract the liquid jet thickness (see Methods and Supplementary Information for a detailed discussion).
% moved to SI:
% Figure~\ref{fig:jet_thickness_interferometric}~a) shows the retrieved jet thickness for each recorded chirped pulse over a time span of 160~seconds, yielding (14~$\pm$~2.4)~$\upmu$m.
% This is in close agreement with the (10~$\pm$~1.1)~$\upmu$m extracted from a commercial thickness sensor (Fig.~\ref{fig:jet_thickness_interferometric}~b), although measured independently under slightly different conditions.
% The latter measurement had been carried out during the experiment's setup phase approximately 24~hours before the timing tool studies commenced.
% Its smaller mean value indicates changed conditions overnight, which also underlines the need to have as much in-situ information as possible during such an experiment, and yields an average value over a much longer measurement period of about 100 ms.
% Since the interference pattern used for the thickness measurement is taken exactly during the pump-probe shot and at the very same lateral position as the time-resolved laser-pump/X-ray probe measurement, the presented data (Fig.~\ref{fig:jet_thickness_interferometric}~a) represents hereby the most accurate values for the real liquid jet thickness during such a pump-probe study.
% original:
%In turn, this information can be useful in additional \textit{a posteriori} corrections of the longitudinal dimension of the interacting sample volume. 
% new:
Since the interference pattern used for the thickness measurement is taken exactly during the pump-probe shot and at the very same lateral position as the time-resolved laser-pump/X-ray probe measurement, this information can be useful in additional \textit{a posteriori} corrections of the longitudinal dimension of the interacting sample volume.

%
%%  end of file

%% file: main-conclusions.tex
% !TeX root = ./main.tex
%
%%  discussion
%
%   original text from the template: the Discussion should be succinct and must not contain subheadings.

We have implemented a novel scheme to measure the relative arrival time between hard X-ray pulses from an FEL and optical laser pulses directly on the sample that was simultaneously under investigation in a time-resolved X-ray pump/optical probe experiment.
We determined a timing accuracy of 22 fs at worst, i.e. for the lowest X-ray pulse energies.
This is sufficient to reliably correct the different arrival times in comparison to the inherent timing jitter between the FEL and optical laser pulses at the SACLA facility under the circumstances of our experiment.
The correlation with the existing SACLA timing tool is excellent, while both methods yield arrival time distributions (timing jitter) of around 500~fs FWHM, even in a measurement where we attenuated the X-ray pulse intensities by a factor of ten.
The small difference in the observed arrival time distributions can be explained by additional jitter sources between both optical laser paths to the two arrival time monitors, which are located a few metres apart from each other.
Therefore, we expect a more reliable timing precision, when measured directly on the sample of interest as in our case.
In a second measurement it was demonstrated that an additional intense optical laser pulse can serve as a sample stimulus in a liquid-phase structural dynamics experiment, while the interferometric encoding approach still delivers accurate in-situ time arrival information.
A full description of the involved electron cascading and propagation of the optical pulses, and details of the estimated change of refractive index upon X-ray and potentially simultaneous intense optical irradiance is subject of a future study.
In order to evaluate the ultimate performance of our tool one would need to compare the corrected timing data in a time-delay scan with a time-dependent X-ray probe signal with all timing tools available.
In an experiment, based on photo-emission of an electron, one would measure a sharp rise after time zero, which would allow to compare the arrival time-corrected rise times with each tool individually, with the goal to identify the superior arrival time monitoring by the shorter rise time.
Furthermore, such a study would then yield detailed data about the remaining timing jitter for timing diagnostics located further away from the actual experiment.
Finally, the analysis of the interference pattern in our arrival time signal additionally allows to derive the thickness of the liquid sheet jet for every laser pulse, potentially allowing for additional correction of a pump-probe signal due to thickness fluctuations in case of sample delivery by a liquid jet. 
The utilisation of a free-flowing liquid jet for in-situ arrival time measurement promises applicability in liquid chemistry experiments at MHz repetition rate XFEL facilities.
%
%%  end of file

%% file: main-methods.tex
% !TeX root = ./main.tex
%
%%  Methods section
%
%   original note from the template: Topical subheadings are allowed. Authors must ensure that their Methods section includes adequate experimental and characterisation data necessary for others in the field to reproduce their work.

%%  fit functions
%
\subsection*{Fit functions}
The fit function used to determine the leading and trailing edge positions of the spectrum measured with the interferometric timing tool is a superposition of a Gaussian error function (describing the rising and falling steps) and a sine function (describing the interference between front and back surfaces) as a function of the detector pixel $x$ (describing the wavelength of the recorded light):
\begin{equation}
    \label{eqn:fit}
    f(x) = { F\left(1 + \text{erf}\left(\frac{\text{x}-\mu}{\sigma}\right)\right) + A\,  \text{sin}\left(\nu(x-\phi) \right)} + C.
\end{equation}
The fitting parameters for the Gaussian error function are its amplitude $F$, centre of the step $\mu$ and width of the step $\sigma$.
It thus describes the interferometric signal onset and cut-off.
The position $\mu$ of the half-rise value of the error function fit is used to define the arrival time of the X-ray pulse.
\newline
The sine function describes the oscillations caused by thin film interference from the front and back surfaces by the supercontinuum traversing the sample.
% new:
These oscillations can be exploited to determine the jet's thickness in-situ and for every single shot (see Supplementary Information).
% moved to SI:
% Its fitting parameters are the sine wave amplitude $A$, the sine wave frequency $\nu$ and its phase $\phi$.
% The thickness of the liquid sheet is then extracted by the analysis of the observed interference patterns near the sharp edges:
% By counting the clearly visible fringes near the interferometric edge positions (usually the first $\pm$2), the thickness can be calculated with
% \begin{equation}
%     \label{eqn:thicknessretrieval}
%     d = \frac{(m_2-m_1)\lambda_1\lambda_2}{2(n_2\lambda_2 - n_1\lambda_1)}
% \end{equation}
% where ($m_2-m_1$) are the number of fringes counted (e.g., the interference maxima) at the % corresponding wavelengths $\lambda_{1,2}$ using their associated refractive indices $n$($\lambda_{1,2}$) = $n_{1,2}$.

%%  Pearson correlation
%
\subsection*{Pearson correlation}
The correlation between two measurements $X$ and $Y$ can be quantified by the Pearson correlation $\rho$ (refs.\cite{Spiegel:book:2017:Statistics,Freedman:book:2007:Statistics}):
\begin{equation}
    \label{eqn:pearsoncorrelation}
    \rho_{X,Y} = \frac{\text{cov}(X,Y)}{\sigma_X \sigma_Y}    
\end{equation}
where cov$(X,Y)$ is the covariance of the two measurements and $\sigma_X$ and $\sigma_Y$ their standard deviation.
A Pearson correlation of $\rho_{X,Y} = 1$ corresponds to a perfect correlation and a value of 0 to a completely uncorrelated measurement.
We use this relation to quantify the correlation between the interferometric and the spatial timing tools in a direct comparison study.

%%  liquid jet
%
\subsection*{Free-flowing flat-sheet liquid jet characterization}
The free-flowing liquid jet system (Microliquids GmbH, Göttingen) produces a thin, flat sheet of liquid based on the collision of two round jets under an angle of 30$^\circ$ to 50$^\circ$, each with selected diameters of a few tens of micrometres.
Adjustments of the backing pressure, the selected nozzle diameter, the relative position and angle of each nozzle, each influences the shape, thickness, flatness and stability of the sheet, allowing a precise control as a function of the viscosity of the liquid.
With this, adjustable flat sheet thicknesses from about 2~$\upmu$m to 300~$\upmu$m are realised.
We used two glass nozzles with 100~$\upmu$m diameter orifices and a volumetric flow rate of up to 80~ml/min. (corresponding to flat sheet speeds up to 60~m/s, see below) using a commercial HPLC pump (Shimadzu LC-20AP), resulting in a leaf-shaped area of the flat sheet of about 0.5$\times$5~mm${}^2$ (horizontal $\times$ vertical dimensions) with a thickness of below 20~$\upmu$m (Supplementary Figure S1). 
This type of liquid jet had been originally developed for soft X-ray spectroscopy experiments\cite{Ekimova:StructDyn:2015:FlatSheetJet}, including appropriate vacuum environments (approx. 10${}^{-3}$~mbar).
However, we optimised the setup for ambient conditions or helium atmospheres as typically used at hard X-ray instruments.
In particular, due to the high volumetric flow rate, a catcher assembly enables the recycling of the sample solution with a minimum total sample volume of approximately 50~ml.

%% file: main-figures.tex
% !TeX root = ./main.tex
%
%%  figure 1: technical drawing and principle of operation
%
\begin{figure*}[p!]
    \centering \includegraphics[width=\linewidth]{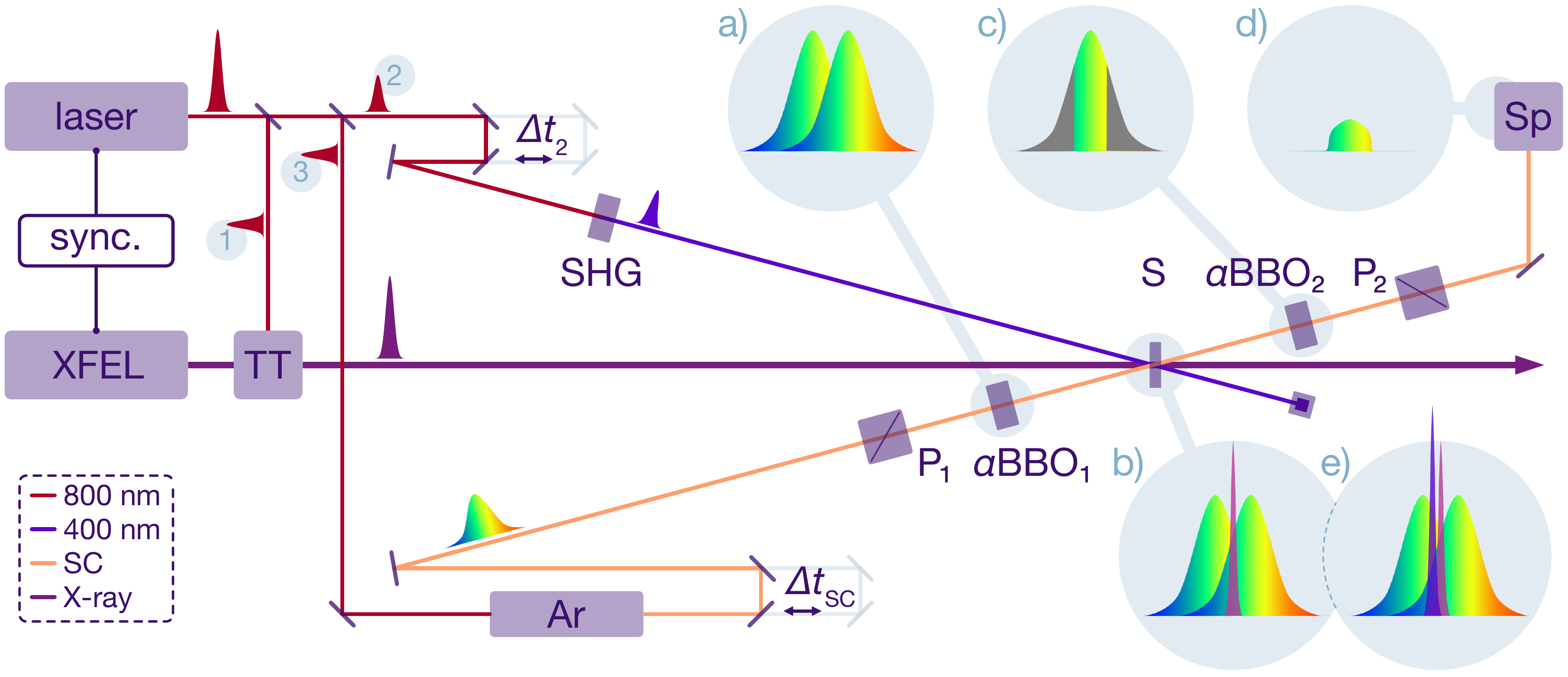}
    \caption{Setup of the XFEL benchmark experiment and principle of operation of the interferometric time arrival monitor.
    The 800~nm beam of the SACLA's synchronised optical laser is split into three branches: the first beam (1) is guided to the facility's standard timing tool TT, the second beam (2) is frequency-doubled (second-harmonic generation, SHG) to a wavelength of 400~nm and serves both for temporal calibration of the interferometric timing tool without X-ray pulses, via the optical delay line adjusting $\Delta t_2$ and for sample excitation with X-ray probing pulses.
    The third beam (3) is used for the generation of a chirped supercontinuum in a pressurised argon cell (Ar), and traverses another optical delay line $\Delta t_\text{SC}$ to arrive with the X-ray pulse on the sample.
    Before the sample, a precision-mounted polariser (P${}_1$) and a 4~mm thick strongly birefringent a-cut $\upalpha$-barium borate crystal ($\upalpha$BBO${}_1$) generates two orthogonally polarised and 970~fs temporally shifted pulses (a).
    This pulse pair is then spatially overlapped with the X-ray pulse in the sample (S), and by adjusting $\Delta t_\text{SC}$, the X-ray pulse arrives temporally in between both supercontinuum pulses (b).
    Inside a second, identical barium borate crystal ($\upalpha$BBO$_2$) with its optical axis rotated by 90${}^\circ$ (c), both supercontinuum pulses are then temporally overlapped again, but still exhibiting their orthogonal polarisation.
    The grey regions indicate those wavelength regions, which experience the same refractive index in the sample, where the leading edge experiences the ground state refractive index while the X-ray induced refractive index change affects the trailing edge of both pulses.
    Subsequent passing through the second polariser (P${}_2$) projects parts of the orthogonal polarised beams onto the same polarisation axis, which then can interfere destructively (d) in the spectrometer (Sp).
    Focusing and collimation of the supercontinuum pulses is realised with curved mirrors to avoid unwanted effects caused by chromatic aberration in lenses.
    Overlapping and additional intense 400~nm optical pulse the actual sample can be excited and for carrying out a traditional optical pump/X-ray probe experiment, while simultaneously recording the mutual optical/X-ray arrival times in-situ (e).}
    \label{fig:interferometric_setup}
\end{figure*}

%%  figure 2: pixel-to-time calibration
%
\begin{figure*}[p!]
    \centering \includegraphics[width=\linewidth]{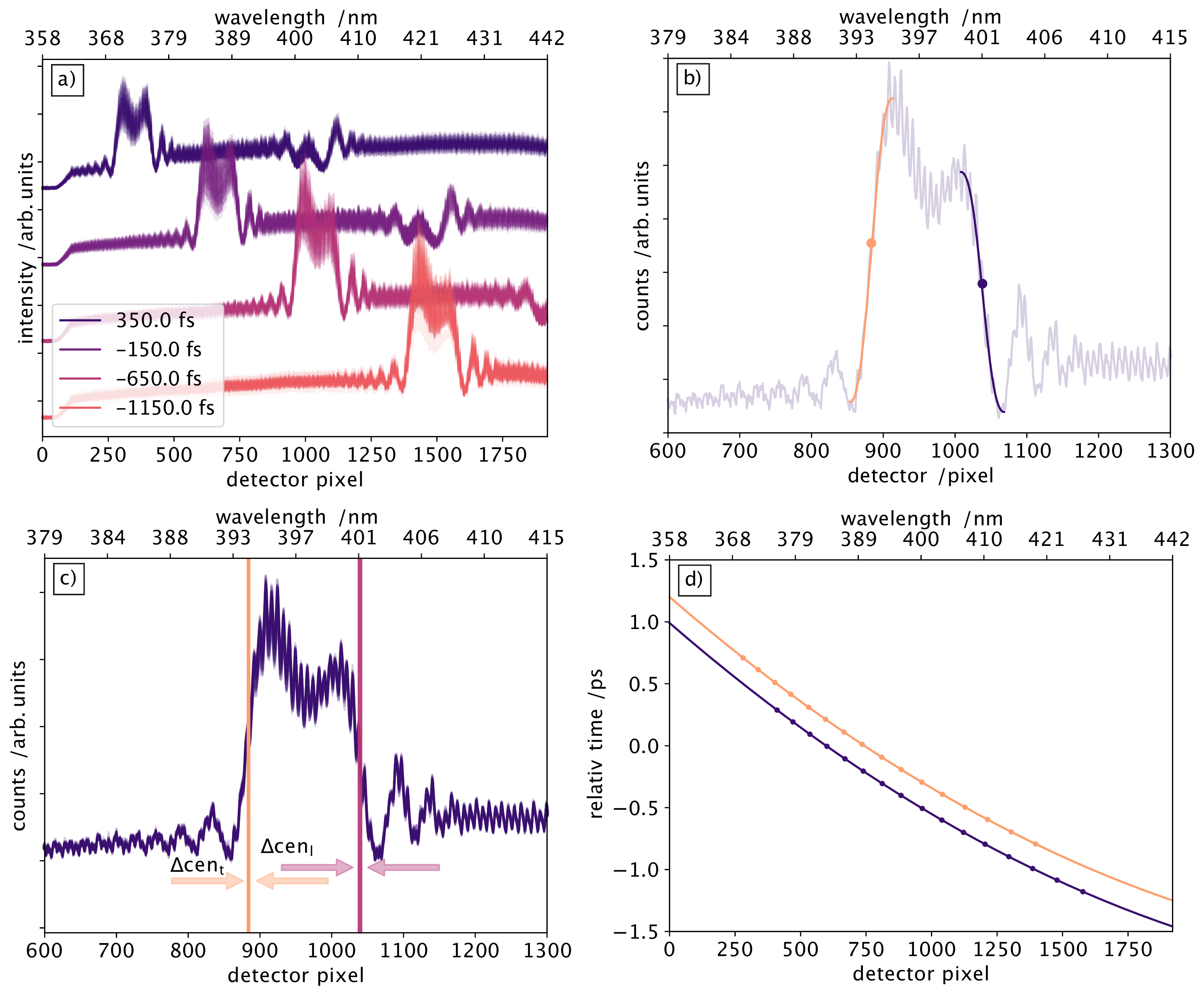}
    \caption{Calibration of the interferometric arrival time measurement setup, utilising a 90~$\upmu$m thick SiO${}_\text{2}$ plate and laser pulses at a wavelength of 400~nm from the same laser system as used for the supercontinuum generation. 
    At each time delay of the 400~nm and the supercontinuum pulses 50 spectra are recorded, which move from higher to lower wavelengths for increasing time delay (a).
    Both edges of every recorded spectrum are then fitted using Eq.~(\ref{eqn:fit}), with one example shown in b) with leading (blue) and trailing (orange) edge fits.
    The centre of the error function is indicated by the dot.
    The 50 individual pump-probe events produce an edge position within a standard deviation of 8~pixels as indicated by the vertical rectangles in c).
    The mapping of detector pixels and thus wavelength to femtosecond time delay (d) results in a calibration constant of 1.8 pixel/fs in the very red part and 1 pixel/fs in the blue part of the recorded spectral range.}
    \label{fig:time_calibration_2dimg}
\end{figure*}

%%  figure 3: raw data of both arrival time detectors and corresponding fits
%
\begin{figure*}[p!]
    \centering \includegraphics[width=\linewidth]{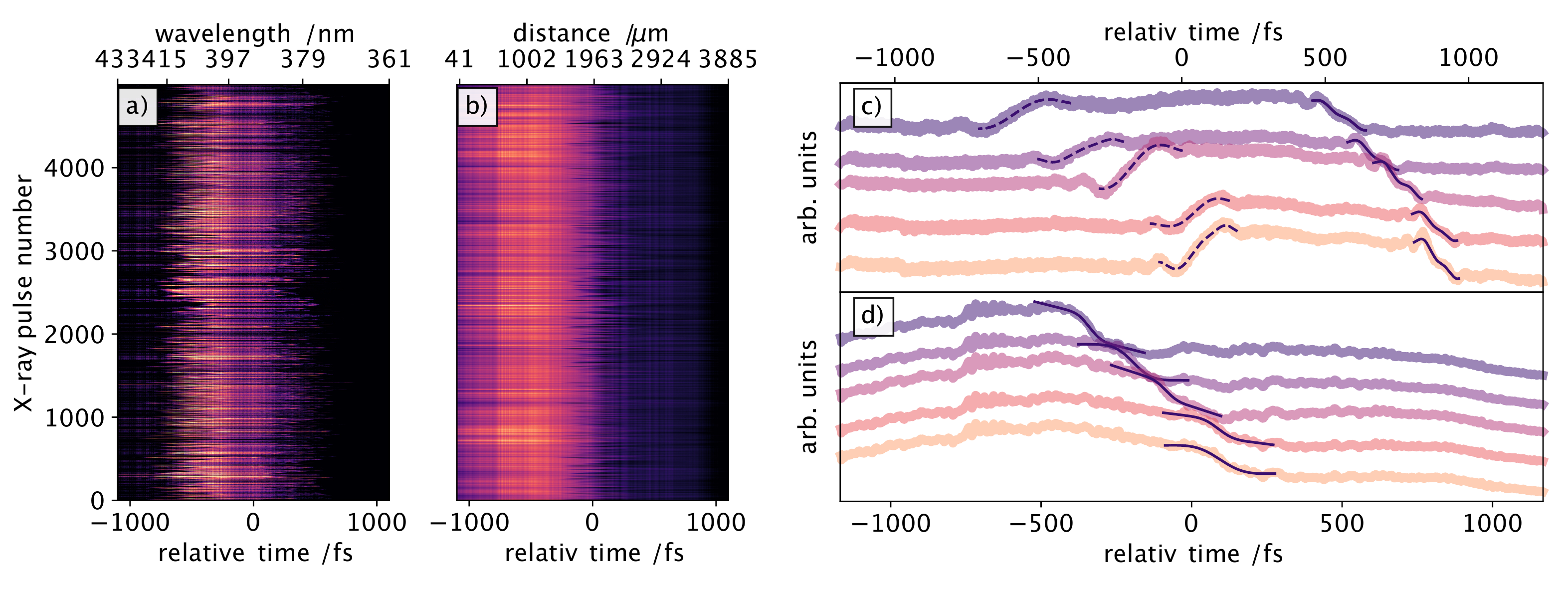}
    \caption{Raw data of the interferometric timing tool a) and SACLA spatial encoding timing tool b), recorded for 5000 X-ray pulses at low fluence.
    Five consecutive X-ray arrival time measurements and corresponding fits for the interferometric timing tool c) and SACLA timing tool d) are shown, displaying the same sequence of events from top to bottom traces, revealing the same overall trend.}
    \label{fig:run550234_raw}
\end{figure*}

%%  figure 4: correlation analysis between the two tools
%
\newpage
\begin{figure*}[p!]
    \centering \includegraphics[width=.75\linewidth]{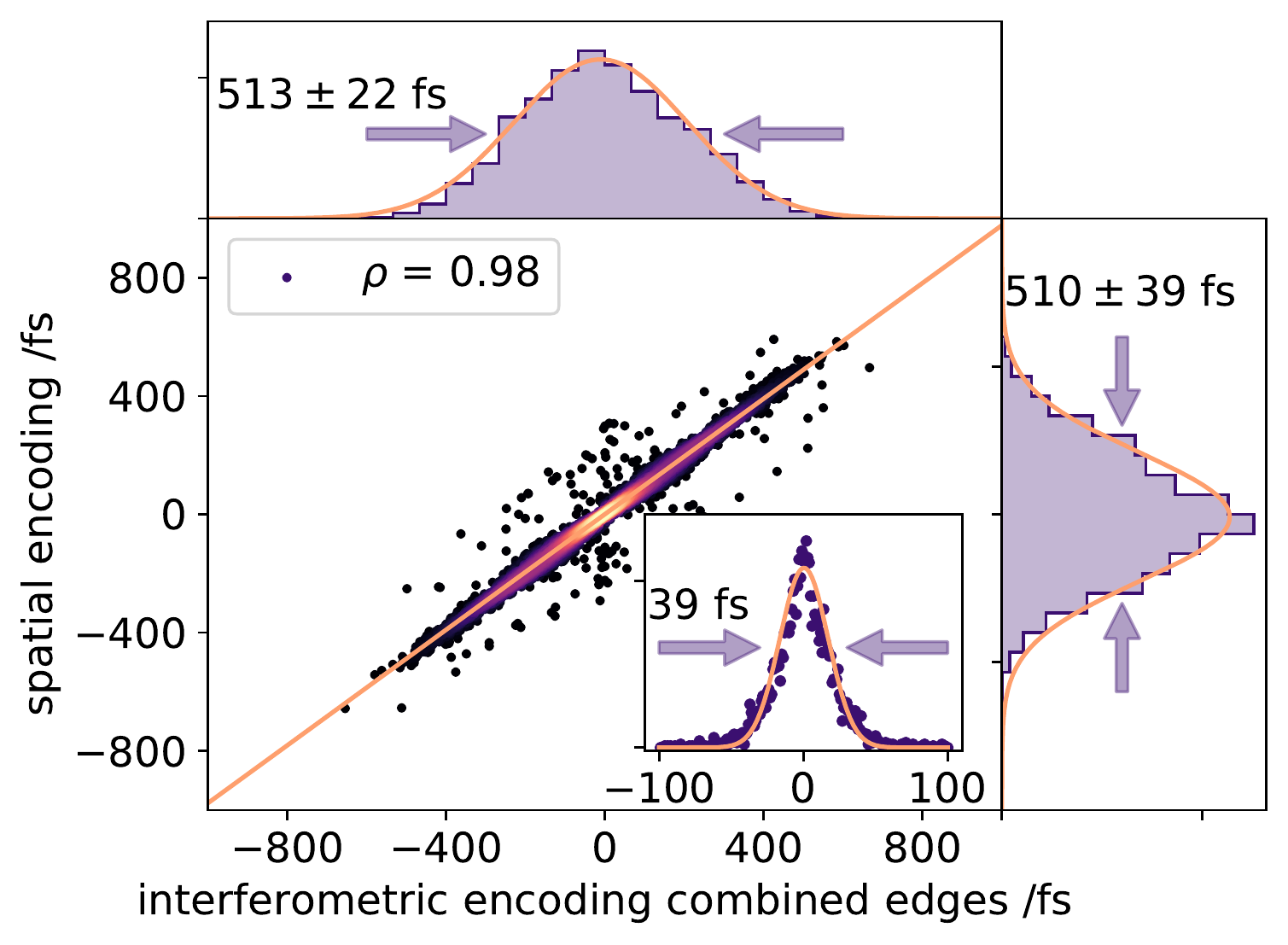}
    \caption{Correlation between interferometric and SACLA spatial encoding timing tools.
    The arrival times retrieved from the interferometric encoding are plotted on the x-axis while the arrival time data of SACLA spatial encoding is plotted on the y-axis.
    Above and right of the main axes the distributions of the interferometric data and the spatial encoded data is shown.
    %The inset shows the residual arrival time jitter between the two timing tools on a logarithmic scale.
    The inset shows the residual arrival time jitter between the two timing tools.}
    \label{fig:run550234_ana}
\end{figure*}

%%  figure 5: arrival time data with a superimposed 400 nm pulse
%
\begin{figure*}[p!]
    \centering \includegraphics[width=.75\linewidth]{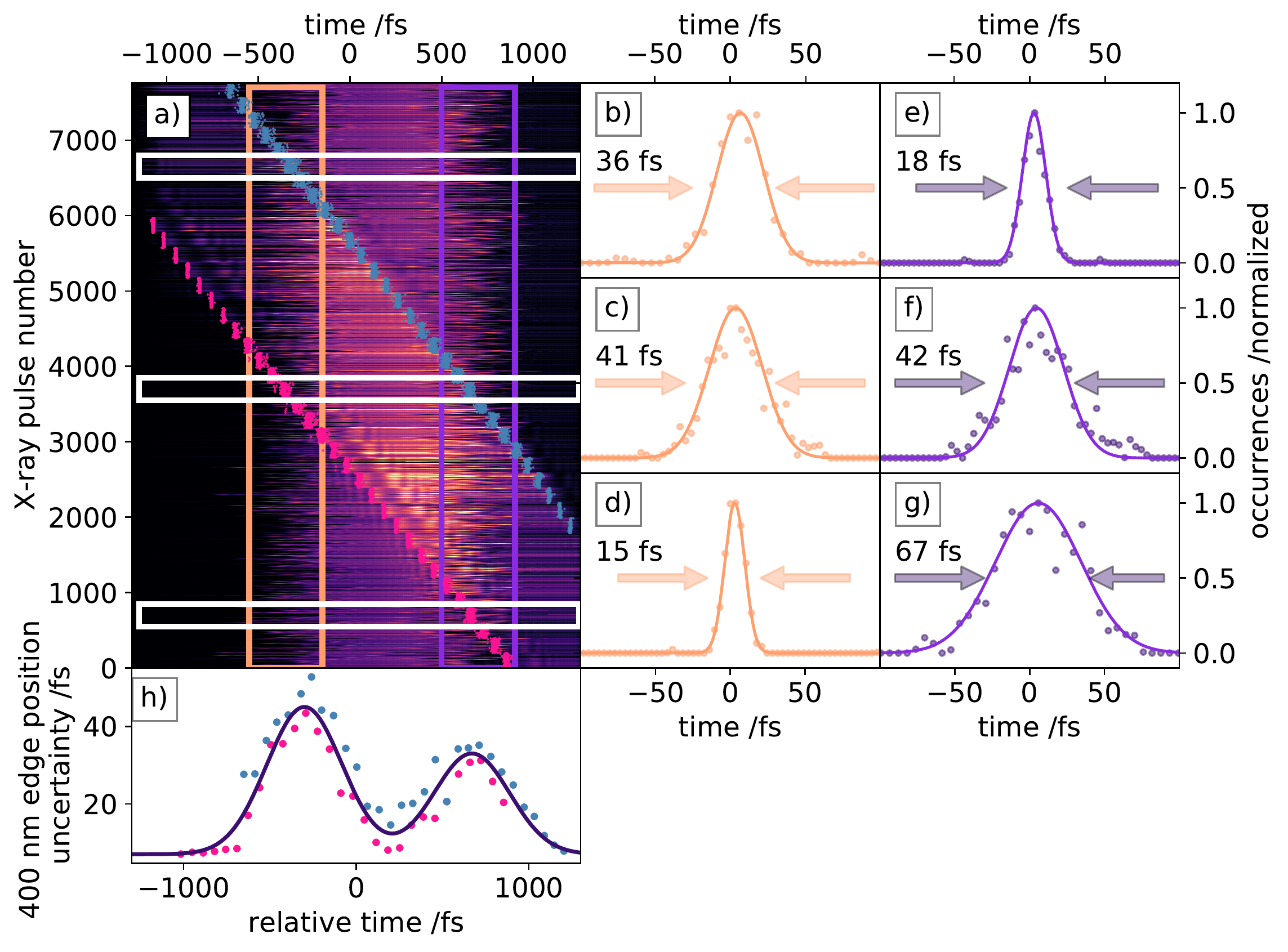}
    \caption{Raw data (a) of the interferometric timing tool with an additional intense 400~nm stimulus for an actual liquid chemistry experiment. The 400~nm pump pulses were delayed from $-$1500~fs to $+$1000~fs with respect to the X-ray arrival times. The additional timing signals induced by the 400~nm stimulus are indicated by the pink (leading edges) and gray-blue (trailing edges) dots. The X-ray induced leading and trailing edge regions are indicated by the orange and purple framed area. Residual timing jitter between the leading (b-d) and trailing (e-g) X-ray induced interferometric edges and the SACLA spatial encoding timing tool at selected time delays. The selected delays are indicated by the white boxes. For b) and e) shots 6600-6800, for c) and f) shots 3600-3800 and for d) and g) shots 600-800 were analysed. The precision with which the 400 nm induced leading (red) and trailing (green) timing edges can be determined is shown in h).}
    \label{fig:run550236_full}
\end{figure*}

% moved to SI:
%%  figure 6: jet thickness determination
%
% \begin{figure*}[p!]
% \centering \includegraphics[width=.75\linewidth]{figures/experiment/jet/both_jet_thickness.pdf}
%     \caption{Thickness of the flat sheet of the liquid jet, extracted from the measured spectral interference pattern alongside the actual timing data in the sample a) and measured prior to the experiment with a commercial confocal device b). In both panels the orange line is indicating a moving average of the data as a guide to the eye.}
%     \label{fig:jet_thickness_interferometric}
% \end{figure*}

%
%% end of file